\documentstyle[11pt]{article}
\begin{document}

\title{Statistical Mechanics Analysis of the 
Continuous Number Partitioning Problem}
\author{ F. F. Ferreira\thanks{Corresponding author. E-mail:
 fagundes@ifsc.sc.usp.br} ~ and  ~ J. F. Fontanari \\
Instituto de F\'{\i}sica de S\~ao Carlos\\
Universidade de S\~ao Paulo\\
Caixa Postal 369\\
13560-970 S\~ao Carlos SP\\
Brazil}
\date{}
\maketitle

\bigskip

\centerline{\large{\bf Abstract}} \bigskip

The number partitioning problem consists of partitioning 
a sequence of positive numbers  $\{ a_1,a_2,\ldots, a_N \}$ into two disjoint sets,
${\cal A}$ and ${\cal B}$,
such that the absolute value of the difference of the sums of $a_j$ over the
two sets is minimized. We use statistical mechanics tools to study analytically
the Linear Programming relaxation of this NP-complete integer programming.
In particular, we calculate the probability distribution of the difference
between the  cardinalities of  ${\cal A}$ and ${\cal B}$
and show that this difference is not self-averaging.

\vspace{1cm}


\vspace{0.3cm} 


\bigskip


\section{Introduction}

Although most of the statistical mechanics analyses of 
stochastic versions of combinatorial optimization problems have focused 
mainly on the calculation of the average cost of the global optima 
\cite{MPV}, the tools of equilibrium
statistical mechanics can also  be used
to evaluate the average performance of simple heuristics 
as well as that of
relaxed versions of the original problem \cite{Jap}.
In this paper we study both numerically and analytically
the Linear Programming (LP) relaxation of a classical NP-complete integer programming
problem, namely, the  number partitioning problem \cite{GJ,KKLO}.

The number partitioning problem (NPP) is stated as follows \cite{KKLO}.
Given a sequence of positive  numbers $\{ a_1,a_2,\ldots,a_N \}$, the 
NPP consists of partitioning them into two disjoint
sets ${\cal {A}}$ and ${\cal {B}}$ such that the difference
\begin{equation}
\left |  \sum_{a_j \in {\cal {A}} } a_j 
- \sum_{a_j \in {\cal {B}} } a_j \right |
\end{equation}
is minimized. Alternatively, we can search for the Ising spin
configurations ${\bf s} =
\left ( s_1,\ldots,s_N \right ) $ that minimize the 
energy or cost function 
\begin{equation}  \label{E_1}
E \left ( {\bf s} \right ) = ~ \left | \sum_{j=1}^N a_j s_j \right |,
\end{equation}
where $s_j = 1$ if $a_j \in {\cal {A}} $ and $s_j = -1$ if 
$a_j \in {\cal {B}} $. 
Also of interest is the problem 
of constrained partitions, in which the difference between 
the cardinalities of sets 
${\cal {A}}$ and $ {\cal {B}}$ is fixed, i.e., 
\begin{equation}  \label{m}
m = \frac{1}{N} ~\left | \sum_{j=1}^N s_j \right | ,
\end{equation}
so that the cardinality of the largest set is $N \left ( 1 + m \right)/2$.
Henceforth we will  restrict  our analysis to the case where the $a_j$'s are 
statistically independent random variables uniformly distributed in the unit interval.

The interest in the NPP stems 
mainly from the remarkable failure of the stochastic heuristic 
simulated annealing  to find good solutions to it, as compared
with the solutions found by deterministic heuristics \cite{JAMS}. 
The reason for that failure is probably due to  
the existence of  order of $2^N$ local minima whose energies are 
of order of $1/N$ \cite{FFF}, which undermines
the usual strategy of exploring the space 
of configurations $\{{\bf s}\}$ through single spin flips.
It is interesting to note that a very simple deterministic 
heuristic, the differencing method
of Karmarkar and Karp, can find with high probability 
solutions whose energies are of
order of $1/N^{\alpha \log N}$ for some $\alpha >0$ \cite{Yakir,ref_kk}. 
For large $N$, however, the energies of the solutions found by the 
differencing method 
are orders of magnitude higher than those predicted by theoretical analyses,
which indicate that the average global optimal energy $\bar{E}_0$ 
is of order of $\sqrt{N}~2^{-N}$ for unconstrained partitions \cite{KKLO,FFF}. 
A recent exact calculation of this quantity yielded $\bar{E}_0 = \sqrt{2 \pi N/3}~2^{-N}$ 
\cite{Mertens}. It must be noted that, in contrast with combinatorial problems
for which the global optimal energy is extensive \cite{MPV}, for the NPP this energy
is not self-averaging \cite{FFF,Mertens} and hence $\bar{E}_0$ cannot 
be viewed as a realization independent minimal energy.

In the LP relaxation we relax the integrality requirement on the 
Ising variables $s_i$ so that they become real variables, i.e., 
$s_i \in \left (-\infty, \infty \right )$. In order to keep these variables
finite we impose a spherical constraint on the norm of the solutions,
\begin{equation}\label{spherical}
\sum_i^N s_i^2 = N .
\end{equation}
Obviously, minimizing the square of the cost (\ref{E_1}) with $s_i$ real but 
constrained
to obey the condition (\ref{spherical}) yields a lower bound to the optimal 
(square) cost of
the corresponding integer programming problem. Moreover, a simple gradient 
descent dynamics
suffices to attain those bounds numerically. In fact, using a Lagrangian multiplier
to handle the constraint (\ref{spherical}) we find that the following dynamics
minimizes the NPP energy,
\begin{equation}\label{dyn_unc}
\frac{\partial s_i}{\partial t } = - \eta \zeta \left ( a_i - \frac{\zeta}{N} s_i \right )
~~~~i=1,\ldots,N
\end{equation}
where $\zeta = \sum_j a_j s_j$ and $\eta$ is 
an arbitrarily small parameter that determines the step-size of the descent.

The remainder of this paper is organized as follows.
In section 2 we show that the LP relaxation yields a trivial lower bound 
(i.e. the LP cost is zero) for unconstrained partitions. That analysis yields,
nonetheless, some interesting pieces of information as, for instance,
the average energy $\bar{E}_c$ obtained by clipping  (i.e. taking the sign of) the spins 
of the global optimal configurations of the LP relaxation. The average performance of
the clipping heuristic is studied in section 3, where it is shown
that $\bar{E}_c$ tends to the average
energy  of a randomly chosen Ising spin configuration, $\bar{E}_c \rightarrow \sqrt{2N/3\pi}$.
In section 4 we calculate
the probability distribution of the difference between the cardinalities of ${\cal A}$
and ${\cal B}$ for 
the LP global optimal configurations,
${\cal P}_c ( m )$, and show that $m$ is not self-averaging. Finally, in section 5 we
present some  concluding remarks.

%
\section{Linear Programming relaxation}
%

In the canonical ensemble formalism of the statistical mechanics
the average value of the optimal energy for
unconstrained partitions is given by
\begin{equation}\label{E_u}
\bar{E}_u =  -
\lim_{T \rightarrow 0} T ~ \left \langle \ln Z_u \right \rangle_a ,
\end{equation}
where $Z_u (T)$ is the partition function
\begin{equation}\label{Z_u}
Z_u (T) = \prod_i \int_{-\infty}^\infty ds_i ~
\delta \left ( N - \sum_j s_j^2  \right ) ~\exp 
\left [ - \frac{E \left ( {\bf s} \right )}{T} \right ]
\end{equation}
with $E \left ( {\bf s} \right )$  given by Eq.\ (\ref{E_1}).
Here  $\delta (x)$ is the Dirac delta and  $T$ 
is the temperature. The notation $\langle \ldots \rangle_a$
stands for the average over the  random variables $a_i$. 
The limit $T \rightarrow 0$ in Eq.\ (\ref{E_u})
ensures that only the states that minimize $E \left ( {\bf s} \right )$
will contribute to $Z_u$.
We now proceed with the explicit evaluation of the partition function
(\ref{Z_u}).
Using the integral representation of the Dirac delta function we write 
\begin{eqnarray}\label{Z_u_1}
Z_u (T)   & = & \int_{-\infty}^\infty 
 \int_{-\infty}^\infty
 \frac{d\zeta d\tilde{\zeta}}{%
2 \pi} \,  \int_{-\infty}^\infty
 \frac{d x}{%
2 \pi} \,\mbox{e}^{i \zeta \tilde{\zeta} + i x N - \mid \zeta \mid /T } \, 
\nonumber \\
  & & \times  \prod_j \int_{-\infty}^\infty  ds_j \,
\exp \left [ -i x s_j^2 - i s_j a_j \tilde{\zeta}  \right ]  .
\end{eqnarray}
The
integrals over $s_j$ and $\tilde{\zeta}$ can easily be performed yielding
\begin{eqnarray} 
Z_u (T)  & = &  \sqrt{\frac{\pi^{N-3}}{4 {\cal M}_2}}
 \int_{-\infty}^{\infty} dx  \, \mbox{e}^{i x N} 
\left ( i x \right )^{(1-N)/2} \nonumber \\
& & \times 
\int_{-\infty}^\infty d\zeta ~\mbox{e}^{- \mid \zeta \mid /T - 
i x \zeta^2/{\cal M}_2}
\end{eqnarray}
where ${\cal M}_2 = \sum_j a_j^2$. At this stage 
the integral over $\zeta$ can be readily carried out by assuming
$T \rightarrow 0$. The result is simply
\begin{equation}\label{Z_u_2}
 Z_u (T)  = T \sqrt{\frac{\pi^{N-3}}{4 {\cal M}_2}}
 \int_{-\infty}^{\infty} dx  \, \left ( i x \right )^{1/2}
\,\mbox{e}^{N G_u (x)} 
\end{equation}
where 
\begin{equation}
G_u (x) = i x + - \frac{1}{2} \, \ln \left ( i x \right ) .
\end{equation}
In the limit of large $N$, the integral over $x$ can be 
evaluated using  the saddle-point method.
Noting that the saddle-point is the imaginary $x_s = 1/2 i $,
the unconstrained partition function is finally written as
\begin{equation}\label{Z_u_f}
 Z_u (T)  = T \sqrt{\frac{1}{2 \pi^2 N  {\cal M}_2}} 
\left ( 1 + \ln 2 \pi \right)^{N/2} .
\end{equation}
Since $Z_u$ decreases  linearly with decreasing $T$, Eq. (\ref{E_u})
yields $\bar{E}_u = 0$. This result was verified numerically
using the gradient descent dynamics (\ref{dyn_unc}) together with an
adaptive prescription to decrease  the step-size  $\eta$ during the 
descent.
%

\section{Clipping heuristic}

%
An easy-to-implement procedure to generate  Ising solutions from the
LP relaxation solutions is to take the sign of the relaxed spins. The
average cost associated to this clipping procedure is given by
\begin{equation}\label{E_c_1}
\bar{E}_c = \lim_{T \rightarrow 0} \left \langle ~ \left \langle ~ 
\left | \sum_j a_j \, \mbox{sign} \, ( s_j ) \right | ~
\right \rangle_T ~ \right \rangle_a
\end{equation}
where $\langle \ldots \rangle_T$ stands for a thermal average taken
with the Gibbs probability distribution, i.e.,
$\exp \left [-E ( {\bf s}) /T \right ]/Z_u$. The zero-temperature limit
ensures that only configurations that minimize the relaxed cost (\ref{E_1})
will contribute to this average. To evaluate Eq.\ (\ref{E_c_1}) we
introduce the auxiliary energy
\begin{equation}
E_{clip} ( {\bf s} ) =  E ( {\bf s} ) + 
h \left | \sum_j a_j \, \mbox{sign} \, ( s_j ) \right |
\end{equation}
with $E  ( {\bf s} ) $ given by Eq.\ (\ref{E_1}). Hence
\begin{equation}
\bar{E}_c = - \lim_{T \rightarrow 0} T ~ \left.
\frac{ \partial \langle \ln Z_{clip} \rangle_a }{\partial h} 
\right |_{h=0} 
\end{equation}
where $Z_{clip}$ is the partition function (\ref{Z_u})  with $E$ replaced
by $E_{clip}$.  Introducing the auxiliary parameter
$\upsilon = \sum_j a_j \, \mbox{sign} \, ( s_j ) $ 
through a Dirac delta function, the calculation of $Z_{clip}$ 
becomes analogous to that presented before, and so we will present
the final results only. We find
\begin{eqnarray}\label{E_c_2}
\bar{E}_c & = & \frac{1}{2\pi} \int_{-\infty}^\infty d \upsilon 
\left | \upsilon \right |  \int_{-\infty}^\infty d \tilde{\upsilon} \, 
\, \mbox{e}^{i \upsilon \tilde{\upsilon} } \left \langle
\prod_j \cos \left ( \tilde{\upsilon} a_j \right ) 
\right \rangle_a \nonumber \\
& =  &  \frac{1}{2\pi} \int_{-\infty}^\infty d \upsilon 
\left | \upsilon \right |  \int_{-\infty}^\infty d \tilde{\upsilon} \, 
\, \mbox{e}^{i \upsilon \tilde{\upsilon} } \, 
\left ( \frac{\sin \tilde{\upsilon}}{\tilde{\upsilon}} \right )^N .
\end{eqnarray}
Assuming that $\bar{E}_c$ does not increase
linearly with $N$, in the thermodynamic limit only the regions close to 
the origin 
($\tilde{\upsilon} = 0$) will contribute to the integral over 
$\tilde{\upsilon}$ in  Eq.\ (\ref{E_c_2}). Hence using 
$\sin \tilde{\upsilon}/\tilde{\upsilon} \approx - \tilde{\upsilon}^2/6$
yields
\begin{equation}\label{E_c_f}
\bar{E}_c = \sqrt{ \frac{2 N}{3 \pi} } .
\end{equation}
We have found a remarkably good agreement between this theoretical
prediction and the properly averaged cost obtained by clipping the 
spherical spins in LP minima  generated by the gradient descent 
dynamics (\ref{dyn_unc}).
Interestingly, for large $N$ the cost (\ref{E_c_f}) is identical to  
the average energy  of a randomly chosen Ising configuration ${\bf s}$,
defined by
\begin{equation}
\bar{E}_r = 2^{-N} ~\prod_i \int_0^1 da_i ~ \sum_{s_i =\pm 1}
\left | \sum_i a_i s_i \right | ,
\end{equation}
which thus demonstrate the complete failure of the clipping heuristic.

%

\section{Probability distribution of cardinalities}

%
As the distinct  LP global minima will have, in general, 
different cardinalities, in this section we calculate analytically
the  probability distribution of the cardinalities difference 
defined by
\begin{equation}\label{P_m_1}
 {\cal P}_c ( m  ) = 
\lim_{T \rightarrow 0} N \left \langle ~ \left \langle ~ 
\delta \left ( ~N  m  - \left | \sum_j s_j \right |~ \right )
\right \rangle_T ~ \right \rangle_a .
\end{equation}
Using the definition of the thermal average this equation is rewritten
as 
\begin{equation}\label{P_c_1}
 {\cal P}_c ( m  ) = 
\lim_{T \rightarrow 0} N  \langle ~ 
\frac{Z_{ m }}{Z_u}
 ~  \rangle_a ,
\end{equation}
where
\begin{equation}\label{Z_m_1}
Z_{ m } = \prod_i \int_{-\infty}^\infty ds_i ~
\delta \left ( N - \sum_j s_j^2  \right ) ~\exp 
\left [ - \frac{E \left ( {\bf s} \right )}{T} \right ]
\delta \left ( ~N  m  - \left | \sum_j s_j \right |~ \right )
\end{equation}
and $Z_u$ is given by Eq.\ (\ref{Z_u}).  The calculation of 
$Z_{ m }$ is a little more complicated  than that  of 
$Z_u$, as it involves the evaluation of an additional integral
due to the extra delta function. However, since the steps are
essentially the same in both calculations we will present the
final result only. In the limits $T \rightarrow 0$ and
$N \rightarrow \infty$ we find
\begin{equation}\label{Z_m_f}
 Z_{ m }   = \frac{T}{N} \,  \sqrt{\frac{1}{\pi^3 {\cal V}}} 
\left ( 1 + \ln 2 \pi \right)^{N/2} 
\exp \left ( - \frac{N {\cal M}_2}{2 {\cal V}}  m^2
\right ) ,
\end{equation}
where ${\cal V} = \sum_j a_j^2 -  N \left ( \frac{1}{N}\sum_j a_j \right )^2$. 
In the limit $N \rightarrow \infty$ we use the self-averaging property,
\begin{equation}
 \frac{1}{N}\sum_i f(a_i) =  \int_0^1 da \, f(a)
\end{equation}
for any function $f$, to write ${\cal M}_2/N= 1/3$ and ${\cal V}/N = 1/12$ so that 
Eq.\ (\ref{P_c_1}) becomes
\begin{equation}
{\cal P}_c ( m ) = \sqrt{\frac{8N}{\pi}} \mbox{e}^{-2 N m^2 } ~~~~m \geq 0.
\end{equation}
Hence, the mean is $\langle  m \rangle = 1/\sqrt{2 N \pi}$ and the
variance, $\sigma^2_m = \left (\pi - 2 \right )/4\pi N$. An important
quantity is the ratio $r_m = \sqrt{\sigma_m^2}/ \langle m \rangle$,
whose vanishing  determines the self-averageness of the random 
variable $m$.
In figures $ 1(a)$ and $1(b)$ we present the results of numerical
experiments to estimate the dependence on $N$ of $\langle m \rangle $ 
and $r_m$, respectively, for three types of configurations: $(i)$
 the global minima of the original NPP obtained through
the exhaustive search in  the Ising configuration  space for $N \leq 26$;
$(ii)$ the legal, Ising configurations obtained with the differencing
method (we refer the reader to ref. \cite{ref_kk}
for a clear presentation of this heuristic); and
$(iii)$ the global minima of the LP relaxation obtained with the
dynamics (\ref{dyn_unc}). We note the very good agreement between the latter
estimate and the analytical predictions. In all cases, the mean 
$\langle m \rangle$  decreases like $N^{-1/2}$ as $N$ increases, while $r_m$ tends
to a nonzero value  ($r_m = \sqrt{\pi/2 -1} \approx
0.755$ for the LP relaxation), indicating that $m$  is not self-averaging 
even in the large $N$ limit, i.e.,
the values of $m$ associated to the configurations under study
depend on the specific realization of the set of random variables
$\{a_j \}$.


\section{Conclusion}


In this paper we have illustrated the usefulness of  
equilibrium statistical mechanics tools to investigate analytically the
average performance of standard relaxation procedures to generate lower bounds
to integer programming problems, as well as to characterize specific properties 
of the minima. The failures of the LP relaxation and the clipping heuristic to 
produce relevant results for the NPP yield
additional evidence to the extreme difficulty of devising heuristics to find
near-optimal solutions to that problem.

\bigskip

\noindent {\bf Acknowledgments}

\bigskip

We thank P.\ Moscato for useful discussions and suggestions.
The work of JFF was supported in part by Conselho Nacional
de Desenvolvimento Cient\'{\i}fico e Tecnol\'ogico (CNPq). FFF
is supported by FAPESP.



\newpage 

\section*{Figure caption}

\bigskip

\parindent=0pt 

{\bf Fig.\ 1} $(a)$ Average cardinalities difference as a function of 
$1/N^{1/2}$ and $(b)$ ratio between the standard deviation
and the average cardinalities difference as a function of $N$.
The convention is $\bigcirc$ (LP relaxation), 
$\bigtriangledown$ (exhaustive search) and $\ast$ (differencing method).
The solid curves are the theoretical predictions for the LP relaxation.


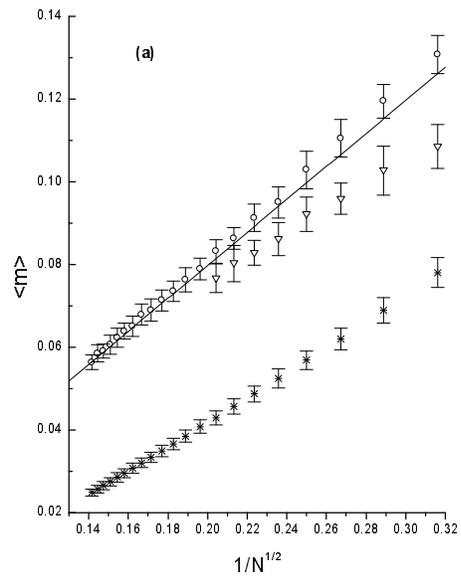
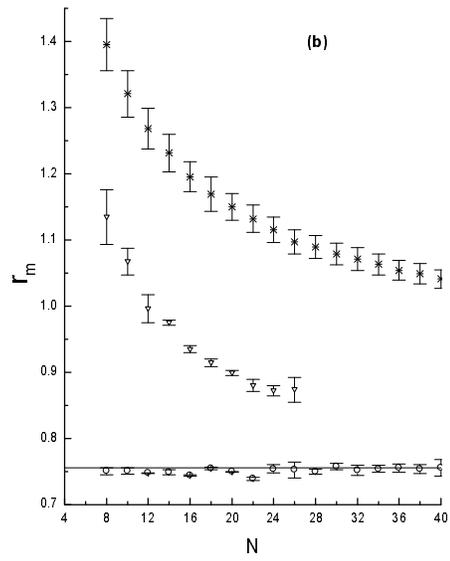
\end{document}